\documentclass[twoside]{article}
\usepackage{fleqn,espcrc2.my,epsfig}

\newcommand{\sv}{{\scriptscriptstyle V}}
\def\llb{\left\{\vphantom{\int}\right.}
\def\rrb{\left.\vphantom{\int}\right\}}
\def\half{{\scriptstyle\frac12\,}}
\def\vec#1{\mbox{\boldmath$#1$}}
\def\({\left(} \def\){\right)}  
\def\lk{\,\left[ \,} \def\rk{\,\right] \,}
   
\def\be{ \begin{equation} }  \def\bea{ \begin{eqnarray} }
\def\ee{ \end{equation} }    \def\eea{ \end{eqnarray} }
\def\Av#1{\left\langle \vphantom{A^b_\mu} #1 \right\rangle}
\newcommand{\msb}{{\overline{{\rm MS}}}}
\newcommand{\msbb}{{{}\over{\rm MS}}}

\def\a{\alpha}
\def\b{\beta}

\def\m{\mu}

\def\p{\pi}

\def\z{\zeta}

\def\pl#1#2#3{Phys.~Lett.~{\bf B {#1}} (19{#2}) #3}
\def\np#1#2#3{Nucl.~Phys.~{\bf B {#1}} (19{#2}) #3}
\def\prl#1#2#3{Phys.~Rev.~Lett.~{\bf #1} (19{#2}) #3}
\def\pr#1#2#3{Phys.~Rev.~{\bf D {#1}} (19{#2}) #3}

\newcommand{\AmS}{{\protect\the\textfont2
  A\kern-.1667em\lower.5ex\hbox{M}\kern-.125emS}}

\title{The Two-Loop Static Potential}

\author{Y. Schr\"oder\address{Deutsches Elektronen-Synchrotron 
        DESY,\\ 
        \hspace*{0.7mm} 22603 Hamburg, Germany}%
        \thanks{Invited talk given at the QCD '99 Euroconference, 
                Montpellier, July 1999.}}

\begin{document}

\begin{abstract}
In quantum chromodynamics (QCD), the binding energy 
of an infinitely heavy quark--antiquark pair in a 
color singlet state can be calculated as a function
of the distance. 
We investigate this static potential of 
QCD perturbatively  and
calculate the full two-loop coefficient, correcting
an earlier result. 
Beyond this order, the perturbative
expansion breaks down.
\end{abstract}

\maketitle

\section{Introduction}

The static potential of quantum chromodynamics (QCD) is 
subject to theoretical investigations since
more than twenty years. Being the non--abelian
analogue of the well--known Coulomb potential of quantum
electrodynamics (QED), this interaction energy of an 
infinitely heavy quark--antiquark pair is a fundamental 
concept which is expected to play a key role in the
understanding of quark confinement. Moreover, the static
potential is a major ingredient in the description of
non--relativistically bound systems like quarkonia,
and it is of importance in many other areas, such as 
quark mass definitions and quark production at threshold.

It is expected that the static potential consists of 
two terms: a Coulomb--like term at short distances, which 
is calculable with perturbative methods, and a long--distance
term responsible for confinement. Even though a perturbative
analysis is not suited to give the full potential, such
a calculation proves very useful. The short--distance part 
of the potential can be utilized as a refined starting point
for the construction of potential models (which have been rather
successful in the past for the description of quarkonia), or
it could describe very heavy systems (like $\bar tt$) to  
good accuracy. Furthermore, it can be compared to the results
of numerical calculations in lattice gauge theory. 
It is natural to define the QCD coupling constant with help
of the potential as $V(r)=-\frac43\frac{\a_\sv(1/r)}r$, the
so-called V--scheme,
using a physical quantity in contrast to the usual coupling
definition in the $\msb$ scheme. In lattice calculations
$\a_\sv$ is regarded as as the 'better' expansion 
parameter \cite{latbetter}. 
For these reasons, and to get a more precise determination of 
$\a_\msbb$ from the lattice, the relation between
the two couplings has to be known.

A first determination of the static potential in (massless) QCD
has been performed by L.~Susskind in the context of
a lecture about lattice gauge theory \cite{Su}. In order to
demonstrate asymptotic freedom in Yang--Mills theory, he
calculated the one-loop pole terms using a Wilson--loop formula
for the potential, and re-derived the first coefficient
of the renormalization group Beta function. 
This work was extended by other groups quite soon, who then
added fermionic contributions \cite{Bi} and two-loop pole terms
\cite{Fi} to the potential, as well as examined the structure
of higher--order corrections qualitatively \cite{ADM}.
Recently, the perturbative static potential has received new
interest, in particular due to its application in top--quark 
production at threshold \cite{TT}, a process that comes within 
experimental reach in the near future.
A complete two-loop calculation for the static potential
was performed in ref.~\cite{Pe}. 
Such an important result
clearly needs confirmation. This is one of the motivations 
of our work on the two-loop potential \cite{YSlett,YSthes}. 
More recently, the effect of fermion masses was considered on 
the two-loop level in ref.~\cite{Me}.

\section{Definition and Expansion}

The static potential is defined in a manifestly
gauge invariant way via the vacuum expectation
value of a Wilson loop~\cite{Fi,mgi_def},
\bea \label{def_WL}
&&\mbox{\hspace*{-7mm}} V(r) = - \lim_{T\rightarrow\infty} 
\frac1T\, \ln \Av{{\cal W}_\Gamma} \;, \\
&&\mbox{\hspace*{-7mm}} {\cal W}_\Gamma = \widetilde{\rm tr}\, 
{\cal P} \exp\(ig \oint_\Gamma dx_\mu A_\mu\) \;.
\eea
Here, $\Gamma$ is taken as a rectangular loop with time extension 
$T$ and spatial extension $r$. The gauge fields $A_\m$ are 
path-ordered along the loop, while the color trace is normalized 
according to $\widetilde{\rm tr}(..)={\rm tr}(..)/{\rm tr}1\!\!1\,$.

In a perturbative analysis it can be shown that, 
at least to the order needed here, all 
contributions to eq.~(\ref{def_WL}) containing connections to
the spatial components of the gauge fields $A_i(\vec r,\pm T/2)$
vanish in the limit of large time extension $T$.
Hence, the definition can be reduced to 
\be \label{def2}
V_{\rm pert} = - \lim_{T\rightarrow\infty} \frac1T\, \ln
\Av{ \widetilde{\rm tr}\, {\cal T} \exp\(\int_x\! J_\mu^a 
A_\mu^a\) } \;,
\ee
where ${\cal T}$ means time ordering and the static sources 
separated by the distance $r=|\vec r-\vec r\,'|$ are given by
\be
J_\mu^a(x) = ig \, \delta_{\mu 0}\, T^a \lk \delta(\vec x-\vec r) 
-\delta(\vec x-\vec r\,') \rk \;,
\ee
where $T^a$ are the generators in the fundamental representation. 
In the case of QCD the gauge group is $SU(3)$. The results 
will be presented for an arbitrary compact semi-simple Lie group 
with structure constants
defined by the Lie algebra $[T^a,T^b]=if^{abc}T^c$. 
The Casimir operators of the fundamental and adjoint representation
are $T^aT^a=C_F$ and $f^{acd}f^{bcd}=C_A\delta^{ab}$. 
$tr(T^aT^b)=T_F\delta^{ab}$ is the trace normalization, 
while $n_f$ denotes the number of massless quarks. 

Expanding the expression in eq.~(\ref{def2}) perturbatively, one 
encounters in addition to the usual Feynman rules the source--gluon 
vertex $ig \delta_{\mu 0} T^a$, with an additional minus sign for the
antisource. Furthermore, the time--ordering prescription
generates step functions, which can be viewed as source
propagators, analogous to the heavy--quark effective theory
(HQET).  

\begin{figure}[t]
\psfig{file=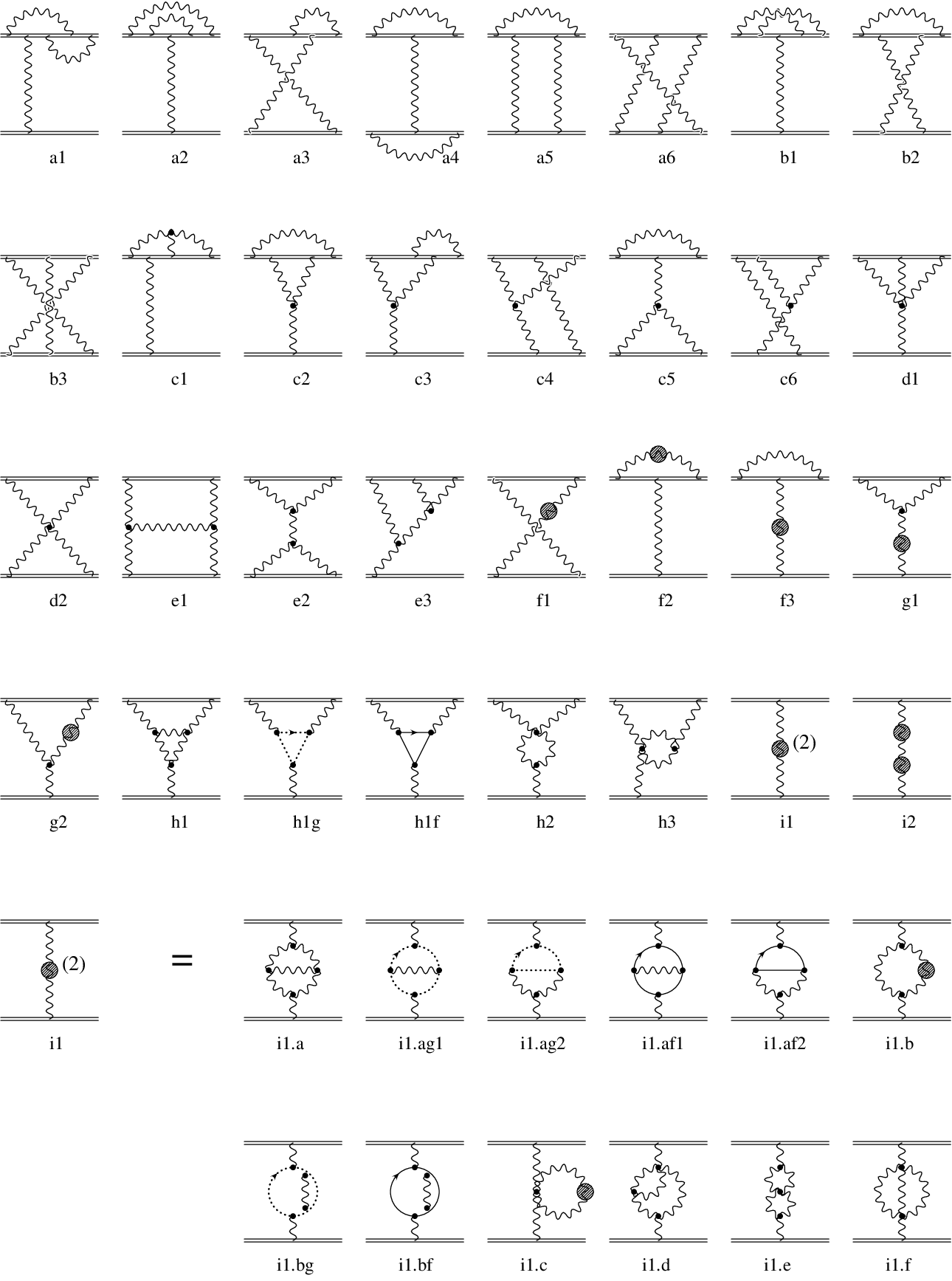,width=7.5cm}
\caption{\label{fig1} Classes of two--loop diagrams 
contributing to the static potential.
Double, wiggly, dotted and solid lines denote source, gluon, ghost 
and (light) fermion propagators, respectively. A blob on a 
gluon line stands for one--loop self--energy corrections. }
\end{figure}

Concerning the generation of the complete set of Feynman diagrams
contributing to the two--loop static potential, there are some
subtleties connected with the logarithm in the 
definition~(\ref{def2}).
All this is explained in detail in \cite{Fi,Pe,YSthes}, so we
only list the relevant exchange diagrams here (see fig.\ref{fig1}). 

\section{Method}

Turning to the purely technical side of the work 
documented here, only part of the Feynman diagrams to be
considered, namely the pure self-energy contributions to the
static potential, are amenable to standard calculation methods
\cite{ChTk,Davy,Ta1,Ta2}.
For the others, essentially being two--point functions also,
but involving non--covariant propagators, the standard methods
need to be generalized. We have developed a general strategy
to deal with these expressions, which is based on a purely algebraic 
reduction to a minimal set of integrals \cite{YSthes}.

The method employed can be briefly 
summarized as follows:
\begin{itemize}
\item Working in momentum space, 
all dimensionally regulated (tensor-) integrals 
are reduced to pure propagator
integrals by a generalization of the method 
of $T$\/--operators \cite{Ta1}.
The resulting expressions are then mapped to
a minimal set of five scalar integrals by means
of recurrence relations, again generalizing \cite{Ta1}
as well as \cite{ChTk} to the case including
static (non-covariant) propagators.
These two steps have been implemented into a FORM package.
Thus, we constructed our method to be complementary to 
the calculation in ref.~\cite{Pe}, assuring a truly independent
check of the result presented therein. 
At this stage, one obtains analytic coefficient functions 
(depending on the generic space--time dimension $D$ as well
as on the color factors and the bare coupling), multiplying 
each of the basic integrals. 
\item The basic scalar integrals are then solved analytically.
Expanding the result around $D=4$ (which is done in both
MAPLE and Mathematica 
considering the complexity
of the expressions), renormalizing and Fourier transforming
back to coordinate space, one obtains
the final results presented below. 
\end{itemize}
 
\noindent Important checks of the calculation include
\begin{itemize}
\item gauge independence of appropriate classes of diagrams, 
\item confirmation of cancelation of infrared divergences,
\item correct renormalization properties.
\end{itemize}

\section{Result and Discussion}

We obtain, as our final result for the two-loop static potential 
in coordinate space,
\bea
 V(r) = -C_F\,\frac{\a_\sv(1/r^2)}{r} \;, 
\eea
with
\bea \label{avseries}
 &&\mbox{\hspace*{-7mm}} \a_\sv(1/r^2) \;=\; \a_\msbb \llb 1 
  +\frac{\a_\msbb}{4\p}\,(a_1+\b_0 L) 
 \nonumber\\&&\mbox{\hspace*{-7mm}} {}
  +\frac{\a_\msbb^2}{16\p^2} \( a_2 +\b_0^2(L^2+\frac{\p^2}{3}) 
  +(\b_1+2\b_0 a_1)L \)
 \nonumber\\&&\mbox{\hspace*{-7mm}} {}
  +{\cal O}(\a_\msbb^3) \rrb \;,
\eea
where $L=2\ln(\bar\m r e^\gamma)$ and $\a_\msbb=\a_\msbb(\bar\m^2)$.
The first two terms of the beta function are given by
$\b_0 = \frac{11}3\,C_A - \frac43\,T_Fn_f$
and
$\b_1 = \frac{34}3\,C_A^2 - 4C_FT_Fn_f -\frac{20}3\,C_AT_Fn_f$.
The one- and two-loop constants read
\bea
  a_1 &=& \frac{31}9\,C_A - \frac{20}9\,T_Fn_f \;, \\
  a_2 &=& \left( \frac{4343}{162} + 4\p^2 - \frac{\p^4}4 + 
  \frac{22}3 \z(3) \right) C_A^2 \nonumber\\
  &&{}- \left( \frac{55}3 - 16\z(3)
  \right) C_FT_Fn_f + \frac{400}{81}\,T_F^2n_f^2 
 \nonumber\\
  &&{}- \left( \frac{1798}{81} + 
  \frac{56}3\,\z(3) \right) C_AT_Fn_f 
 \;, \label{resulT}
\eea
respectively.
As it has to be, the coefficients prove to be gauge  
independent.
Comparing our two-loop result for $a_2$ with \cite{Pe}, we find
a discrepancy of $2\p^2$ in the pure Yang--Mills term 
($\propto C_A^2$).
This amounts to a $30\%$ decrease of $a_2$ for the case of $n_f=0$,
and a $50\%$ decrease for $n_f=5$ (for $SU(3)$), which is the case 
needed for $t\bar t$ threshold investigations. 
This difference can be traced back to a specific set of diagrams,
and it is clarified with the author of ref.~\cite{Pe}~\footnote{We 
thank M.~Peter for checking this result.}.

There are now numerous concepts for  a 'renormalization group
improvement', i.e. for an 'optimal'  choice of the $\msb$ scale
parameter $\bar\m$ in order to reduce large logarithmic
corrections. Examples include the 'natural choice' 
$\bar\m=e^{-\gamma}/r$
and the choice $\bar\m=\exp(-\gamma-a_1/2\b_0)/r$, which eliminates
the one-loop coefficient completely. Due to this freedom, it is not
very illuminating to present plots of the coordinate space
potential. A general feature is that, at increasing distance, 
the large two-loop coefficient begins to dominate quite soon, even
causing the potential to decrease again above some $r_{\rm crit}$,
to signal that the perturbative approach can be followed up to
this critical distance at most. For a discussion of a variety
of scale choices we refer to ref.~\cite{Pe}. The smaller
coefficient $a_2$ found in our calculation does not change the
plots presented there qualitatively. The reason is that the term
$\frac{\p^2}{3}\b_0^2$, which shows up as a result of the Fourier
transform, is of the same order as $a_2$.

Concerning the convergence of the series (\ref{avseries}), let us 
give some numbers. For SU(3), $T_F=\half$ and the 'natural' scale 
choice, we have
\bea
 \a_\sv &=& \a_\msbb \llb 
  1 +\frac{\a_\msbb}{\p}\,(2.6-0.3n_f) 
 \nonumber\\&&{} 
  +\frac{\a_\msbb^2}{\p^2}\,(53.4-7.2n_f+0.2n_f^2) +.. \rrb \;.
\eea
In ref.~\cite{Pe}, the first number in the two-loop term was $64.5$.
Apparently the convergence
properties do still not look very promising.

A comparison with four--dimensional quenched lattice results
is given in ref.~\cite{Balinew}. There, it is concluded
that the perturbative potential already fails to describe the 
slope of the lattice potential at the smallest distances that
are numerically tractable, hence invalidating the possibility
to match the two potentials at small distances.

Summarizing, we have re-calculated the two--loop static potential
by a method complementary to the approach in \cite{Pe}.
We have developed an algorithm which enables us to
work in general covariant gauges throughout. Confirming the
fermionic contributions to the two--loop coefficient $a_2$,
we find a substantial deviation in the pure gluonic part of $a_2$.
The source of the discrepancy could be identified.
The bad convergence of the perturbative series does not improve
considerably taking into account the new value of $a_2$. Hence, 
the use of a physical coupling, defined by the potential, as 
expansion parameter seems to be disfavored. Further studies 
are clearly needed to clarify the role of higher--order corrections. 
There has always been discussion about whether the perturbative 
Wilson--loop formula is a good definition of 
the static potential, which 
can be questioned due to possible infrared divergences in
higher orders \cite{ADM}. A redefinition was proposed
recently \cite{BPSV}, which becomes effective at the three-loop
level.

\section*{Acknowledgments}

We would like to thank W.~Buchm\"uller, M.~Spira, T.~Teubner and 
M.~Peter for valuable discussions and correspondence, respectively.

\end{document}